# Modeling Dispositional and Initial learned Trust in Automated Vehicles with Predictability and Explainability

[1]Jackie Ayoub, [2]X. Jessie Yang, [1]Feng Zhou

[1]Department of Industrial and Manufacturing Systems Engineering, University of Michigan-Dearborn, Dearborn, MI, USA

[2]Department of Industrial and Operations Engineering, University of Michigan, Ann Arbor, Ann Arbor, MI, USA

Corresponding author:

Feng Zhou, 4901 Evergreen Road, Dearborn, MI 48128, Email: fezhou@umich.edu

**ABSTRACT**

Technological advances in the automotive industry are bringing automated driving closer to road use. However, one of the most important factors affecting public acceptance of automated vehicles (AVs) is the public's trust in AVs. Many factors can influence people's trust, including perception of risks and benefits, feelings, and knowledge of AVs. This study aims to use these factors to predict people's dispositional and initial learned trust in AVs using a survey study conducted with 1175 participants. For each participant, 23 features were extracted from the survey questions to capture his/her knowledge, perception, experience, behavioral assessment, and feelings about AVs. These features were then used as input to train an eXtreme Gradient Boosting (XGBoost) model to predict trust in AVs. With the help of SHapley Additive exPlanations (SHAP), we were able to interpret the trust predictions of XGBoost to further improve the explainability of the XGBoost model. Compared to traditional regression models and black-box machine learning models, our findings show that this approach was powerful in providing a high level of explainability and predictability of trust in AVs, simultaneously.

# 1. Introduction

Automated vehicles (AVs) are the next technological revolution due to their advantages (e.g., safer, cleaner, and more efficient) compared to manual driving (Elrod, 2014). Although the automated driving systems will one day handle the whole driving task and allow drivers to do non-driving related tasks (SAE, 2018), the public seems reluctant to adopt the technology. A survey study showed that only 37% of their participants would probably buy an AV (J. D. Power, 2012). Menon (2015) showed that 61.5% of Americans were not willing to use AVs. Such results may be partially due to Uber's and Tesla's crashes involving automated driving, which have shaken consumers' trust in AVs. A recent AAA study (Edmonds, 2019) revealed that three out of four Americans were afraid of using AVs. Therefore, Shariff et al. (2017) and Bansal et al. (2016) concluded that the key barriers to the adoption of AVs are psychological rather than technological, and the most critical one is probably trust.

According to Lee and See (2004), trust is defined as "*the attitude that an agent will help achieve an individual's goals in a situation characterized by uncertainty and vulnerability*". Researchers identified many factors affecting people's trust in AVs. Ayoub et al. (2019) summarized the factors affecting trust into three categories, including 1) human-related factors (i.e., culture, age, gender, experience, workload, and knowledge about AVs), 2) automation-related factors (i.e., reliability, uncertainty, and user interface), and 3) environmental-related factors (i.e., risk, reputation of original equipment manufacturers). Hoff and Bashir (2015) identified three layers of variability in human-automation trust, including dispositional trust, situational trust, and learned trust. Dispositional trust indicates people's enduring tendency to trust, situational trust measures the construct related to trust dynamics during human-automation interaction in specific contexts, while learned trust is related to how past experiences in automated systems influence individuals' current level of trust in automation.

Estimating trust in AVs is challenging, especially when the majority of the public does not have much interaction experience with AVs. Raue et al. (2019) suggested that people's experience in manual driving should potentially shape their trust assessment in AVs. Along the same line, Abe et al. (2017) made use of manual driving characteristics (e.g., speeds and time headway) to investigate driver's trust in automated driving in terms of overtaking and passing patterns. Researchers have used linear and logistic regression models (Raue et al., 2019) and other machine learning (e.g., Support Vector Machines (SVMs)) methods to investigate the effects of various factors on trust in AVs (López and Maag, 2015; Liu et al., 2011). While regression models are limited to predict people's trust in AVs, machine learning models, such as SVMs, are capable of providing better predictability at the price of their explainability, i.e., they are considered as black boxes (Rudin, 2019; Adadi and Berrada, 2018).

To fill the research gaps, this study proposes a method that provides both good predictability and explainability of trust in AVs, using eXtreme Gradient Boosting (XGBoost) (Chen and Guestrin, 2016) and SHapley Additive exPlanations (SHAP) (Lundberg et al., 2020). By trust in AVs, we mainly measure dispositional and initial learned trust prior interacting with an AV as in (Hoff and Bashir, 2015), because we collected the data through a survey without providing chances for the participants to interact with AVs. First, XGBoost is an ensemble machine learning model based on a large number of decision trees that use an optimized gradient boosting system. It has the advantage to perform parallel processing, to approximate greedy search, and to improve the learning process in the smallest amount of time without overfitting. It was demonstrated that XGBoost had the best prediction performance and processing time compared to gradient boosting and random forest (Chen and Guestrin, 2016). Second, in order to improve the explainability of the XGBoost model, we used SHAP (Lundberg et al., 2020), which uses a game theoretic approach that explains the output of a machine learning model. It combines optimal credit allocation with local and global explanations using the classic Shapley values from game theory and their related extensions. In summary, this study made the following contributions: (1) We proposed a machine learning model that estimates people's dispositional and initial learned trust in AVs with good predictability and explainability; (2) We identified critical factors affecting people's dispositional and initial learned trust in AVs; (3) We identified the main effects and the interaction effects between the critical factors that explain trust prediction.

## 2. Related Work

**2.1. Factors Affecting Trust in AVs**

To increase the public usage of AVs, it is essential to understand the factors affecting people's trust perception. Many researchers have consistently reported the effects of risks, benefits, knowledge, and feelings on trust (Walker et al., 2016; Raue et al., 2019; Rudin-Brown and Parker, 2004; Parasuraman and Miller, 2004).

**Perception of Risks:** Risk is considered to be an intrinsic aspect affecting trust, i.e., when the perceived risk of a situation is high, a higher level of trust is needed to rely on AV's decisions (Numan, 1998; Kim et al., 2008; Pavlou, 2003). Therefore, it is essential to consider factors associated with risks in AVs when evaluating trust (Rajaonah et al., 2008). Zmud et al. (2016) reported that safety risks due to system failures were the major concerns of using AVs. Moreover, Menon et al. (2016) showed that one third of US drivers were worried about the risks of misusing their private AV data. Li et al. (2019) demonstrated that the perceived risks and trust in an AV were affected by introductory information related to system reliability. Therefore, it is important to include risk perception and an appropriate level of information regarding AVs to evaluate trust in the early stages of driver-vehicle interactions.

**Perception of Benefits:** Many researchers have found that the perception of benefits is related to improving trust in AVs, which subsequently leads to user acceptance and adoption (Choi and Ji, 2015; Bearth and Siegrist, 2016). One of the major benefits associated with AVs is to reduce vehicle crashes and to save lives. Vehicle crashes lead to injury of 2.2 million Americans each year (NHTSA, 2010) and the cost associated with these crashes is around $300 billion (Bearth and Siegrist, 2016). Therefore, the safety enhancement behind AVs should be focused on creating crash-less vehicles (Johnson, 2012; Fagnant and Kockelman, 2015; Paden et al., 2016). As a matter of fact, human factors were reported to be the cause of 90% of crashes and the death of over 30 thousand Americans per year (Elrod, 2014). AVs are accurate and quicker to react in case of emergency since they can optimize the decision before taking any actions. Aside from

improving safety, AVs can bring other social benefits, including reducing congestions, fuel consumption, and CO2 emission (Fagnant and Kockelman, 2015), and so on.

**Knowledge about AVs:** Another important factor influencing trust is the knowledge of the public regarding the capabilities and limitations of AVs. A lack of knowledge in automation leads to mistrust or over-trust of the true capabilities of the system (Parasuraman and Riley, 1997). Doney et al. (1998) presented a direct effect of knowledge on trust, where knowledge reduced uncertainty which in return increased trust. Khastgir et al. (2018) demonstrated that providing introductory knowledge about AVs to the participants increased their level of trust in the system. To calibrate trust, the authors suggested the concept of information safety to ensure safe interaction with AVs. Holmes (1991) argued that trust developed with the accumulation of knowledge from increasingly more experience from the past. Therefore, experience plays an important role in shaping our trust assessment. For instance, Ruijten et al. (2018) demonstrated that mimicking human behavior using intelligent user interfaces improved drivers' trust in AVs. Edmonds (2019) showed that participants who had advanced driver-assistance systems (ADAS) in their vehicles were 68% more likely to trust these features than the drivers who did not have them.

**Effect of Feelings:** Trust is composed of two components: a cognitive component and an affective component (Lewicki and Brinsfield, 2011; Cho et al., 2015). The cognitive component is based on judgements, beliefs, competence, stability, and expectations while the affective component is based on positive and negative emotions that shape our trust (Lewis and Weigert, 1985). For example, positive emotions were found to improve takeover performance in AVs, which further led to trust in AVs (Du et al., 2020) while negative emotions, such as concerns and worries, made parents trust automated school buses less (Ayoub et al., 2020). Furthermore, Peters et al. (2006) explained that affect influenced our stored knowledge, which further guided our acceptance and trust. Hence, emotion can be used to evaluate trust. According to Hancock and Nourbakhsh (2019), the majority of drivers had no chance to experience AVs yet. Thus, this inexperience makes it harder to evaluate their trust in the system. Raue et al. (2019) suggested that feelings related to people's experience in driving could shape their perception of risks, benefits, and trust in AVs. Specifically, Baumeister et al. (2001) showed that negative emotions were more significant in shaping judgment than positive ones.

**2.2. Modeling Techniques of Trust in AVs and Automation**

Many researchers used questionnaires (Körber, 2018) and behavioral methods (Miller et al., 2016; Jessup et al., 2019) to evaluate trust in automation and in AVs. For instance, Körber, (2018) built a multidimensional model to measure trust in automation using a survey study. The model was composed of 19 parameters, including reliability, understandability, propensity to trust, familiarity, and intentions. Miller and Perkins (2010) developed a survey to study trust in automation by focusing on 5 components of trust including competence, predictability, dependability, consistency, and confidence. Furthermore, Lee and See (2004) summarized the factors affecting trust in automation into a three-dimensional model, including performance, process, and purpose. Jian et al. (2000) built a scale system to measure trust using an experimental study that explored the similarities and differences between trust and distrust in automation. Raue et al. (2019) used linear regression ($R^2 = 0.72$) to model interests in using AVs and logistic regression ($R^2 = 0.31$) to model parents' attitudes toward children riding in AVs alone. Both models identified significant factors (e.g., risk perception, benefit perception, negative emotions in manual driving) influencing the dependent variables, but no prediction results were reported. Commonly, trust models are modeled using a linear combination of the input factors, which identify significant factors that influence trust in AVs and other automation systems. However, they did not report prediction results. Therefore, machine learning techniques were proposed in modeling trust in AVs. For example, Liu et al. (2011) investigated the usage of two machine learning models: linear discriminant analysis for feature importance and decision trees for classification for large-scale systems (e.g., product recommendation systems, Internet auction sites) with false rates between 10% and 19%. Guo and Yang (2020) developed a personalized trust prediction model based on the Beta distribution and learned its parameters using Bayesian inference. López and Maag (2015) designed a generic trust model capable of processing various trust features with an SVM technique. On their simulated trust dataset, they obtained 96.61% accuracy. Akash et al. (2018) developed an empirical trust model of object detection in AVs and they used a quadratic discriminant classifier and psychophysiological measurements, such as electroencephalography (EEG) and galvanic skin response (GSR). Their model's best accuracy was 78.55%. Such models were able to predict people's trust in AVs to a large extent by aggregating numerous factors. However, the relative importance in predicting trust in AVs tends to be not obvious in such black-box models. Unlike prior work, we propose a research method that combines XGBoost and SHAP to help increase the predictability and explainability of trust in AVs, simultaneously.

**3. System architecture**

The proposed system architecture is illustrated in Fig. 1 with the following steps:

(1) Data Collection: We collected a dataset using an online survey on Amazon Mechanical Turks (AMTs). The survey was developed in Qualtrics and it was integrated in AMT to collect participants' responses.
(2) Data Cleaning: We reviewed the participants' responses and removed invalid data.
(3) XGBoost Model Construction: We used a 10-fold cross validation process to optimize the parameters of XGBoost to train the model.
(4) XGBoost Model Evaluation: To evaluate the performance of the XGBoost model, we compared it with a list of machine learning models using various performance metrics, including accuracy, receiver operator characteristics area under the curve (ROC_AUC), precision, recall, and F1 measure.
(5) SHAP Explanation: To improve the explainability of the XGBoost model, SHAP was used to explain the model predictions both globally and locally.

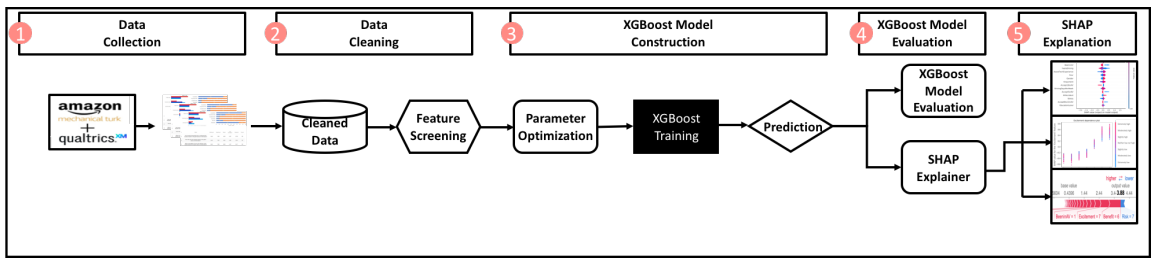

**Fig. 1.** Flow chart of the proposed system architecture to predict Trust.

## 4. Method

### 4.1. Participants and Apparatus

A total number of 1175 participants located in the United States took part in the online survey using AMTs (Seattle, WA, www.mturk.com/). AMT is a web-based survey company, operated by Amazon Web Services, which has recently become popular in fast data collection (Paolacci et al., 2010). The questionnaire was developed in Qualtrics (Provo, UT, www.qualtrics.com), a web-based software to create surveys. Participants who gave nonsensical answers (i.e., unreasonable driving experience compared to their age, using letters instead of numbers to represent the number of driving years, using the same pattern to answer all the questions, and completing the survey too quickly) were excluded from the study. After the screening, we had a total number of 1054 participants (47.5% females, 52.2% males, and 0.3% others). The age distribution and the education distribution of the participants are shown in Table 1. Participants were compensated with $0.2 upon completion of the survey. The study was approved by the Institutional Review Board at the University of Michigan.

**Table 1**. Age and education distribution of the participants in the study

| Age Distribution | <18 | 18-24 | 25-34 | 35-44 | 45-54 | 55-64 | >=65 |
|---|---|---|---|---|---|---|---|
| | 0.1% | 8.3% | 37.7% | 22.7% | 14.4% | 10.9% | 5.9% |
| Education Distribution | Professional degree | Doctoral degree | Master's degree | Bachelor's degree | Some college | Associate degree | High school degree or less |
| | 1.2% | 0.9% | 18.3% | 43.3% | 16.9% | 11.5% | 7.9% |

### 4.2. Survey Design

We investigated various factors associated with AVs, including knowledge, experience, feelings, risk and benefit perceptions, and behavioral assessment to predict trust using a survey study. The survey questions were adapted from (Raue et al., 2019; Jian et al., 2000) and are shown in Table 2. Participants' knowledge about AVs was measured using their eagerness level to adopt a new technology, knowledge level about AVs, and knowledge about AV crashes. Experience questions were related to the experience of using ADAS and the experience of trying AVs. As for Benefit and Risk related questions, participants had to assess how beneficial and risky the AVs were. In regard to the behavioral assessment related questions, participants were asked if they would let a child under 5 years old, between 6 to 12 years old, between 13 to 17 years old, and above 18 years old use an AV alone. Since the majority of the public had no experience in AVs yet, we asked them to rate their feelings (i.e., Control, Excitement, Enjoyment, Stress, Fear, and Nervousness) based on their experience in manual driving. Among all the items in the survey, those related to knowledge and experience directly measured participants' initial learned trust while others measured their dispositional trust. We provided abbreviated names for the survey questions to use them throughout the paper as shown in Table 2.

**4.3. XGBoost Model Construction**

In this study, the XGBoost classifier was selected for predicting trust in AVs (Chen and Guestrin, 2016). The boosting algorithm combines multiple decision trees into a strong ensemble model and reduces the bias by reducing the residual error at each iteration where each decision tree learns from the previous one. This process is done by adjusting the weights of decision trees while iterating the model sequentially. More accurate decision trees are given more weights. XGBoost implements the same boosting technique with an additional regularization term. During the optimization process, an optimal output value for each tree is obtained by iteratively splitting each tree to minimize its objective function.

To build a tree, the process follows the exact greedy algorithm where it starts with all the training examples, and then it calculates the split loss reduction or gain for the root of the tree. Once the gain for all the split trees is calculated, the tree with the maximum gain is considered as the optimal split. The gain value should be positive in order for the selected tree to continue growing. After building the trees, pruning is performed to remove the sections with low effect on the classification. Then, an output value is calculated for each leaf which will be used to make predictions. Using these predictions, the same described process is used to build a second tree. The XGBoost algorithm combines both software and hardware optimization abilities, which result in great performance with less computational resources by performing parallel computing.

**Table 2**. Survey questions, categories, and scale

| Categories | Survey Questions | Abbreviation | Scale |
|---|---|---|---|
| **General** | 1) What is your gender? | Gender | |
| | 2) What is your age? | Age | |
| | 3) What is the highest level of school you have completed or the highest degree you have received? | EducationLevel | |
| | 4) Do you have a valid driving license? | DrivingLicense | |
| | 5) For how many years have you been a driver? | YearsDriving | |
| | 6) On average, how many days a week do you drive? | DrivingDaysPerWeek | |
| **Knowledge** | 7) What is your eagerness level to adopt new technologies? | EagertoAdopt | From 1 (extremely low) to 7 (extremely high) |
| | 8) What is your knowledge level in regard to autonomous vehicles? | KnowledgeinAVs | From 1 (extremely low) to 7 (extremely high) |
| | 9) Have you heard any stories about autonomous vehicles being involved in accidents? | AVAccident | Yes / No |
| **Experience** | 10) Please indicate how much experience you have with vehicle driving assistance technology (for example: cruise control, adaptive cruise control, parking assist, lane keeping assist, blind spot detection, or others) | AssistTechExperience | From 1 (extremely low) to 7 (extremely high) |
| | 11) Have you ever been in an autonomous vehicle? | BeeninAV | Yes / No |
| **Benefit and risk perception** | 12) What is the risk level of using an autonomous vehicle? | Risk | From 1 (extremely low) to 7 (extremely high) |
| | 13) How beneficial it is to use an autonomous vehicle? | Benefit | From 1 (extremely low) to 7 (extremely high) |
| **Behavioral assessment** | 14) Would you let a child who is under 5 years old use an autonomous system alone? | Assess5inAV | |
| | 15) Would you let a child who is between 6 and 12 years old use an autonomous system alone? | Assess6to12inAV | Yes / No |
| | 16) Would you let a child who is between 13 and 17 years old use an autonomous system alone? | Assess13to17inAV | |
| | 17) Would you let an adult who is above 18 years old use an autonomous system alone? | Assess18inAV | |
| **Feelings** | 18) How much do you feel in control (for example: attentive, alert) when you are driving? | Control | |
| | 19) How much do you feel excited when you are driving? | Excitement | |
| | 20) How much do you enjoy driving? | Enjoyment | From 1 (extremely low) to 7 (extremely high) |
| | 21) How much do you feel stressed when you are driving? | Stress | |
| | 22) How much do you feel scared when you are driving? | Fear | |
| | 23) How much do you feel nervous when you are driving? | Nervousness | |
| **Trust** | 24) In general, how much would you trust an autonomous vehicle | Trust | From 1 (extremely low) to 7 (extremely high) |

In this research, we removed the highly correlated predictor variables before starting the training process in XGBoost using the Pearson correlation coefficient. The correlation coefficient was high between age and number of driving years (0.88) and between fear and nervousness (0.87). Therefore, age and nervousness were removed. We defined the response variable as a binary one, (i.e., trust = 1 (extremely high, moderately high, and slightly high), sample size = 624, and distrust = 0 (extremely low, moderately low and slightly low), sample size = 430) by converting its 7-point Likert scale. In the next step, we trained the XGBoost classifier with 10-fold cross validation to optimize the accuracy of the prediction using a randomized search for hyperparameters. The learning objective used in this study was reg: logistic regression. After we constructed the model, we compared XGBoost with other machine learning models using various performance metrics, including accuracy, ROC_AUC, precision, recall, and F1 measure. Accuracy is the fraction of corrected prediction samples divided by the total samples. ROC plots the true positive rate against the false positive rate at various threshold settings, and ROC_AUC measures the performance of a classifier in distinguishing between the two classes. Precision is defined as true positive/(true positive + false positive), recall as true positive/(true positive + false negative), and F1 measure as the harmonic mean of precision and recall, i.e., 2*precision*recall/(precision+recall) (Zhou et al., 2017).

**4.4. Explaining XGBoost Model Using SHAP**

Shapley value is a method from coalitional game theory (Shapley, 1953), in which each player is assigned with payouts depending on their contribution to the total payout when all of them cooperate in a coalition. In our study, in the case of XGBoost model, each feature (i.e., predictor variables in XGBoost) has its fair contribution to the final prediction of trust perception on AVs. Predicting if one participant trusts or distrusts AVs can be considered as a game, and the gain in this game is the actual prediction for this participant minus the average prediction for all the participants' data. For example, if we use three feature-value sets, i.e., Benefit = 7, BeeninAV = 1, and KnowledgeinAVs = 7 to predict trust in AVs, the predicted Trust is 7 and if we use Benefit = 7 and KnowledgeinAVs = 7 to predict trust in AVs, the predicted Trust is 5. Assuming we want to calculate the Sharply value of the feature-value set, BeeninAV = 1, the contribution from the above example is 7 - 5 = 2 in trust prediction. However, this is only one coalition, we need to repeat the same process for all the possible coalitions and obtain the average of all the marginal contributions. Mathematically, the Shapley value of a feature-value set is calculated as follows (Shapley, 1953):

$$\varphi_i(v) = \sum_{S \subseteq N \setminus \{i\}} \frac{|S|!(n-|S|-1)!}{n!} (v(S \cup \{i\}) - v(S)), \quad (1)$$

where $n$ is the total number of features, $S$ is a subset of any coalition of the features $N$, where the summation extends over all subsets $S$ of $N$ that do not contain feature $i$, and $v(S)$ is the contribution of coalition $S$ in predicting trust in our study. The difference between the trust prediction and the average trust prediction is fairly distributed among all the feature-value sets in the data. Therefore, it has a solid theory in explaining machine learning models.

One limitation is that when the number of features increases (so is the exponential number of coalitions), the computation needed will be exponentially expensive. According to game theoretically optimal Shapley values, Lundberg and Lee (2017) and Lundberg et al. (2020) proposed an efficient method to calculate SHAP values, especially for tree-based models, such as XGBoost. Therefore, we can use SHAP to explain XGBoost both globally and locally. Globally, we can study how SHAP values rank the features based on their importance, how SHAP values change with regard to different feature-value sets, and how one feature interacts with another. Locally, we can explain individual predictions. Among them, the interaction effect is defined as the additional combined feature effect minus individual main feature effects:

$$\varphi_{i,j}(v) = \sum_{S \subseteq N \setminus \{i,j\}} \frac{|S|!(n-|S|-2)!}{n!} (v(S \cup \{i,j\}) - v(S \cup \{i\}) - v(S \cup \{j\}) + v(S)), \quad (2)$$

Thus, SHAP can produce an $n$ by $n$ interaction matrix and automatically can identify the strongest interaction effect given one specific feature. In this research, after training the XGBoost model, SHAP was used to explain the model predictions (Lundberg and Lee, 2017) by calculating the importance of each feature, by evaluating the interaction effects between the features globally, and by explaining individual predictions locally.

## 5. Results

### 5.1. Survey Results

We calculated participants' mean responses and the standard deviations as shown in Fig. 2. The knowledge-related questions indicated that the majority of the participants had a relatively high level of knowledge about AVs — 75.1% had a high level of eagerness to adopt a new technology (i.e., by high, we mean a Likert scale value greater than or equal to 5, moderate refers to a Likert scale value of 4, and low refers to a Likert scale value less than or equal to 3), 51% had a high level of knowledge in AVs, and 76.4% of the participants knew about accidents related to AVs. As for the experience related questions, the majority showed a low level of experience in AVs— 46% of the participants had a high level of experience in ADAS and 77.3% had never been in an AV.

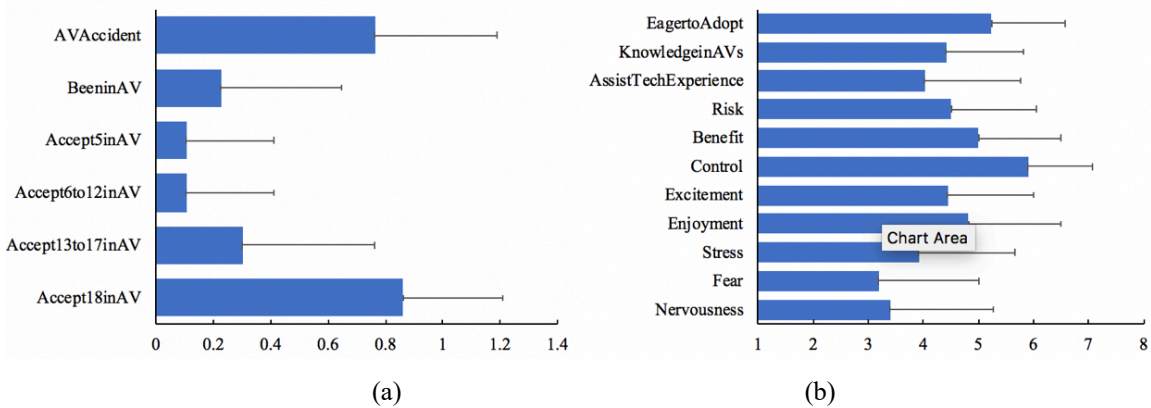

**Fig. 2.** Mean values and standard deviations of the predictor variables. (a) "0" = No, 1 = "Yes"; (b) "1" = Extremely low, "2" = Moderately low, "3" = Slightly low, "4" = Neither low nor high, "5" = Slightly high, "6" = Moderately high, "7" = Extremely high.

Furthermore, the majority considered AVs as beneficial (71%), but risky (57%). In regard to behavioral assessment of AVs, 89% of the participants were reluctant to let a child under 5 or between 6 to 12 use an AV alone and 70% were reluctant to let a child between 13 and 17 use an AV alone. However, 86% were willing to let a child above 18 use an AV alone. Feelings related questions showed that the majority of the participants reported a high level of control (91%) and a high level of excitement (51%) and enjoyment (64%) while driving. In addition, 58% of the participants had a low level of fear and nervousness of driving, but 44% of the participants considered driving as being stressful.

**5.2. XGBoost Performance**

The performance of the XGBoost prediction model, including accuracy, ROC_AUC, precision, recall, and F1 measure, is shown in Table 3 using a 10-fold cross validation strategy. In order to compare the performance of XGBoost with other algorithms (see Table 3), we also performed a 10-fold cross validation strategy. We found that XGBoost performed the best across almost all the metrics (except precision) among the list of the machine learning models including logistic regression, decision trees, naive Bayes, linear SVM, and random forest.

Table 3. Performance measures comparison between different models

| Models | Accuracy | ROC_AUC | Precision | Recall | F1 Measure |
|---|---|---|---|---|---|
| Logistic Regression | 83.1% | 0.90 | 82.1% | 82.6% | 82.3% |
| Decision Tree | 83.5% | 0.87 | **82.9%** | 82.8% | 82.9% |
| Naïve Bayes | 81.6% | 0.90 | 81.2% | 80.8% | 81.0% |
| Linear SVM | 84.4% | 0.91 | 82.8% | 84.3% | 83.5% |
| Random Forest | 83.1% | 0.90 | 81.3% | 83.3% | 82.3% |
| XGBoost | **85.5%** | **0.92** | 82.5% | **91.6%** | **86.8%** |

**5.3. SHAP Global Explanation**

**Importance of Predictor Variables:** To understand the importance of each factor in predicting trust in AVs, we examined SHAP feature (i.e., predictor variable) importance and summary plots. The SHAP feature importance plot sorts the features by the mean of the absolute SHAP value over all the samples i.e., $\frac{1}{M}\sum_{j=1}^{M}|\phi_{ij}(v)|$, where $M$ is the total number of the samples. The SHAP summary plot also combines feature importance with feature effect. Note the unit of the SHAP value here is log odds as the objective function was set as logistic regression in training the XGBoost model. The summary plot lists the most significant factors in a descending order as illustrated in Fig. 3(a). The top factors (e.g., Benefit, Risk, Excitement, KnowledgeinAVs, EagertoAdopt) contributed more to the prediction. To obtain more information about the factors, we also explored the summary plot in Fig. 3(b). Each data point (i.e., each participant) has three characteristics, including 1) the vertical location that shows importance ranking based on the overall SHAP value of a particular predictor factor, 2) the horizontal spread that depicts whether the value has a small or large effect on the prediction, and 3) the color coding that describes the value of the factor from low (i.e., blue) to high (i.e., red) gradually. For instance, a small value of the Benefit factor has shown to reduce the log odds of the prediction of trust by almost 2.5, whereas a large value of the Benefit factor increases the prediction by almost 2. Such results not only show the importance of the predictor variables, but also help us understand how they influence the prediction results. Furthermore, the spread of the important factors tends to be wider than those of the unimportant factors, and the SHAP value of the majority of the unimportant factors tends to be around 0, such as EducationLevel.

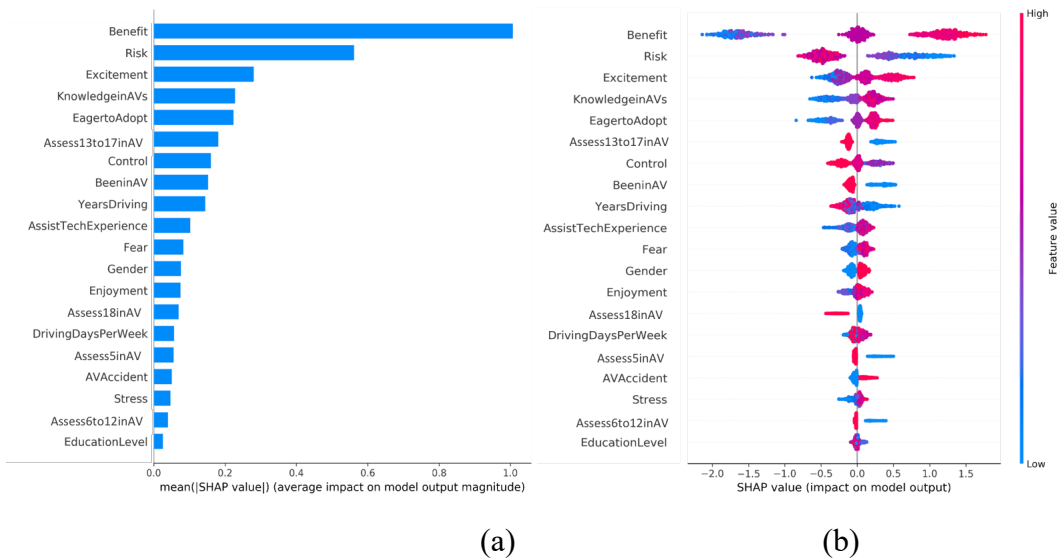

(a)            (b)

**Fig. 3.** (a) SHAP feature importance plots (b) SHAP summary plot.

**Dependence Plot:** To further understand the relationship between the predictor variables and the response variable, we examined their individual SHAP dependence plots which can capture both, the main effects of individual predictor variables and the interaction effects between predictor variables. Fig. 4 shows the SHAP dependence plots of the top five most important factors (i.e., Benefit, Risk, Excitement, KnowledgeinAVs, and EagertoAdopt) and a continuous variable, i.e., YearsDriving. For instance, to understand the impact of Benefit on trust as captured by the XGBoost model, the SHAP dependence plot is shown in Fig. 4(a). The horizontal axis represents the actual values of the Benefit factor from the dataset, and the vertical axis represents the effect of the factor on the prediction. For the main effect, the plot shows an increasing trend between the factor Benefit and the target trust. It also shows the interaction effect between Benefit and BeeninAV automatically selected by the SHAP model. Out of the participants who scored low on benefits of AVs, those who had experience with AVs trusted AVs more than those who had no experience. On the other hand, out of the participants who scored high on benefits of AVs, those who had experience with AVs trusted AVs less than those who had no experience with AVs. The SHAP dependence plot of Risk is illustrated in Fig. 4(b). We can observe that risks in AVs are negatively correlated with trust in AVs. Meanwhile, among the participants who scored low on risks in using AVs, those who had no experience in AVs trusted AVs more than those with experience. On the other hand, among the participants who scored high on risks in AVs, those who had experience with AVs trusted AVs more than those who did not. The effect of Excitement on trust is illustrated in Fig. 4(c). The higher the excitement about manual driving, the higher the likelihood to trust AVs. And among the participants with a low level of excitement about driving, those who scored high on perceived risks in AVs trusted AVs less than those who scored low on perceived risks. However, among the participants with a high level of excitement about driving, those who scored high on perceived risks in AVs trusted AVs more than those who scored low on perceived risks in AVs. Fig. 4(d) illustrates the effect of KnowledgeinAVs on trust. The increasing slope indicates that the more the Knowledge in AVs, the higher the likelihood to trust AVs. For the participants who rated low in knowledge in AVs, those with low perceived risks in AVs trusted AVs more than those with high perceived risks in AVs. However, when the participants rated high in knowledge in AVs, those with high perceived risks in AVs trusted AVs more than those with low perceived risks in AVs. The increasing slope in Fig. 4(e) shows that the more eager the participants are to adopt a new technology, the higher the likelihood is to trust AVs. Out of the participants who were not eager to adopt a new technology, the interaction effect was not clear. However, out of the participants who were eager to adopt a new technology, those being not fearful of driving trusted AVs more than those being fearful of driving. In Fig. 4(f), we see a decreasing slope which illustrates that people with more experience in driving are less likely to trust AVs. For the participants with driving experience between 10 and 40 years, those who reported a high level of perceived benefits trusted AVs more than those who reported a low level of perceived benefits.

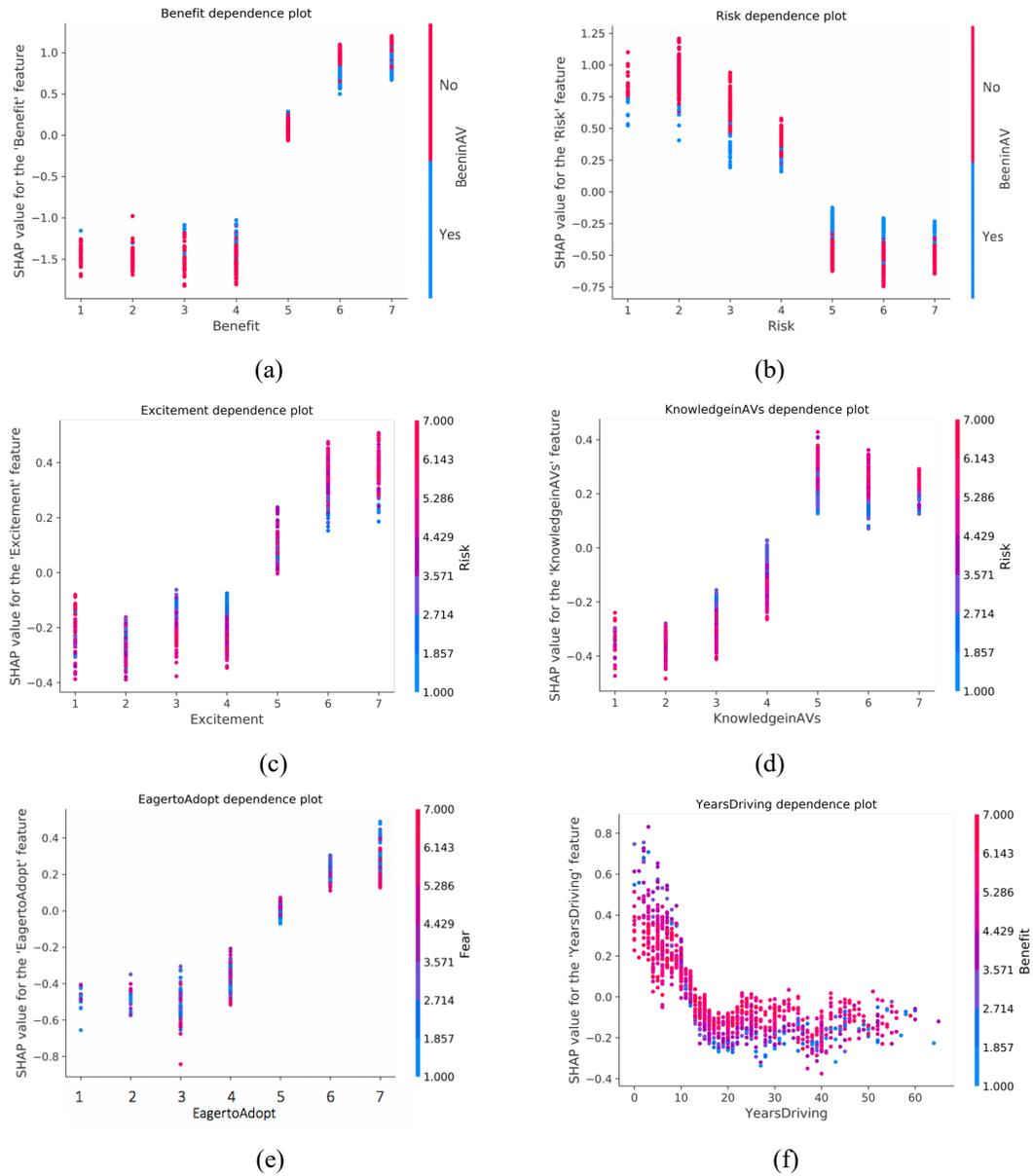

**Fig. 4.** SHAP dependence plots. (a) Benefits, (b) Risk, (c) Excitement, (d) KnowledgeinAVs, (e) EagertoAdopt, and (f) YearsDriving. "1" = Extremely low, "2" = Moderately low, "3" = Slightly low, "4" = Neither low nor high, "5" = Slightly high, "6" = Moderately high, "7" = Extremely high.

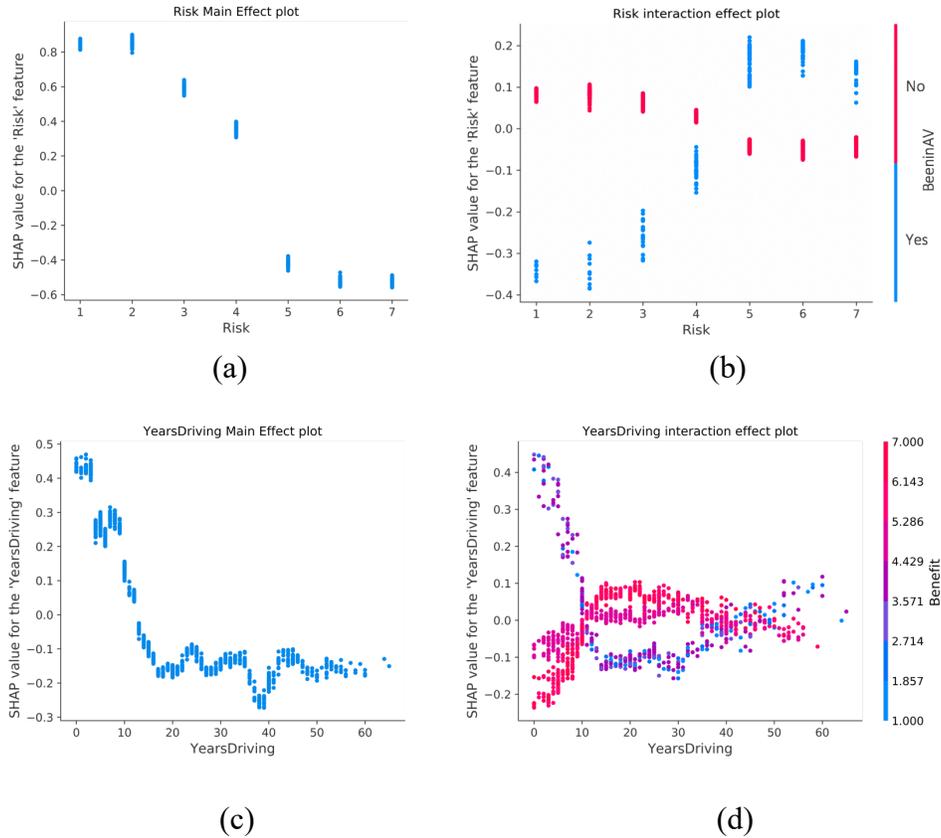

**Fig. 5.** SHAP main effects and interaction effects derived from SHAP dependence plots. "1" = Extremely low, "2" = Moderately low, "3" = Slightly low, "4" = Neither low nor high, "5" = Slightly high, "6" = Moderately high, "7" = Extremely high.

**Main Effects and Interaction Effects:** The SHAP dependence plot has rich information, which incorporates both main effects of individual predictor variables and interaction effects between two predictor variables. The interaction effects are demonstrated by the vertical dispersion as shown in Fig. 4. Such interaction shows the effect of the two predictor variables on the response variable at the same time. We can also separate the main effects and interaction effects in individual plots. Take the Risk SHAP dependence plot in Fig. 4(b) as an example. Its main effect and interaction effect with BeeninAV are shown in Fig. 5(a) and Fig. 5(b). There is little vertical dispersion in the main effect. The interaction effect is also more apparent suggesting that at lower Risk levels, participants who experienced AVs trusted AVs less than those who did not experience AVs. However, at higher Risk levels, participants who experienced AVs trusted AVs more than those who did not experience AVs. Take the YearsDriving as another example. Its main effect and interaction effect with Benefit are shown in Fig. 5(c) and Fig. 5(d). Also, less vertical dispersion is observed in the main effect plot, and the interaction effect tends to be more apparent. That is, only when YearsDriving is larger than 10 and smaller than 40, more Benefits lead to a stronger likelihood to trust AVs.

In Table 4, we presented the sum of the main effects (i.e., $\sum_{j=1}^{M} |\phi_{ij}(v)|$, where $M$ is the total number of the samples) and selected interaction effects of the six predictor variables corresponding to Fig. 4. The larger the magnitudes of the main/interaction effects, the more important they are to predict trust. Furthermore, we also calculated the correlation coefficients between the selected predictor variables and their SHAP values and between the selected predictor variables and the response variable, i.e., trust. Although all the correlations are significant, the correlations with SHAP values are stronger, indicating that XGBoost tends to capture the correlations better than linear models.

**Table 4.** Rich information obtained from SHAP dependent plots for selected predictor variables

| Predictor Variables | Main effect | Selected interaction effect | Correlation with SHAP values | Correlation with Trust |
|---|---|---|---|---|
| Benefit | 945.94 | :BeeninAV: 22.13 | 0.89 | 0.61 |
| Risk | 543.47 | :BeeninAV: 41.67 | -0.90 | -0.37 |
| Excitement | 233.42 | :Risk: 31.39 | 0.86 | 0.25 |
| KnowledgeinAVs | 265.51 | :Risk: 24.51 | 0.87 | 0.41 |
| EagertoAdopt | 234.70 | :Fear: 5.96 | 0.92 | 0.42 |
| YearsDriving | 190.59 | :Benefit: 39.02 | -0.69 | -0.24 |

The *p* values of all the correlation coefficients in the table are smaller than 0.001

## 5.4. SHAP Local Explanations

In order to show how SHAP explains individual cases, we tested it on two randomly selected observations as illustrated in Fig. 6. The plots show the different factors contributing to pushing the output value from the base value which represents the average model output over the training dataset. The base value is defined as the mean prediction value (Lundberg et al., 2018), which is 0.5358 in our case. Factors pushing the SHAP value (i.e., log odds) larger are shown in red while those pushing the SHAP value smaller are shown in blue. In Fig. 6(a), the model produced a large SHAP value in predicting trust which was consistent with the ground truth (i.e., trust) because the participant perceived the AV with a high level of Benefits (i.e., 6), BeeninAV = Yes, a high level of Excitement (i.e., 6), a high level of KnowledgeinAVs (i.e., 7), Assess13to17inAV = Yes, a high level of EagertoAdopt (i.e., 6), YearsDriving (i.e., 4), even though the participant perceived the AV with a high level of Risk (i.e., 7). In Fig. 6(b), the model produced a small SHAP value, which was consistent with the ground truth (i.e., distrust) mainly due to a neutral level of Benefit, a high level of Risk (i.e., 5), a neutral level of EagertoAdopt (i.e., 4), a low level of KnowledgeinAVs (i.e., 2), 21 YearsDriving, a low level of Excitement (i.e., 1), and a low level of Fear (i.e., 1).

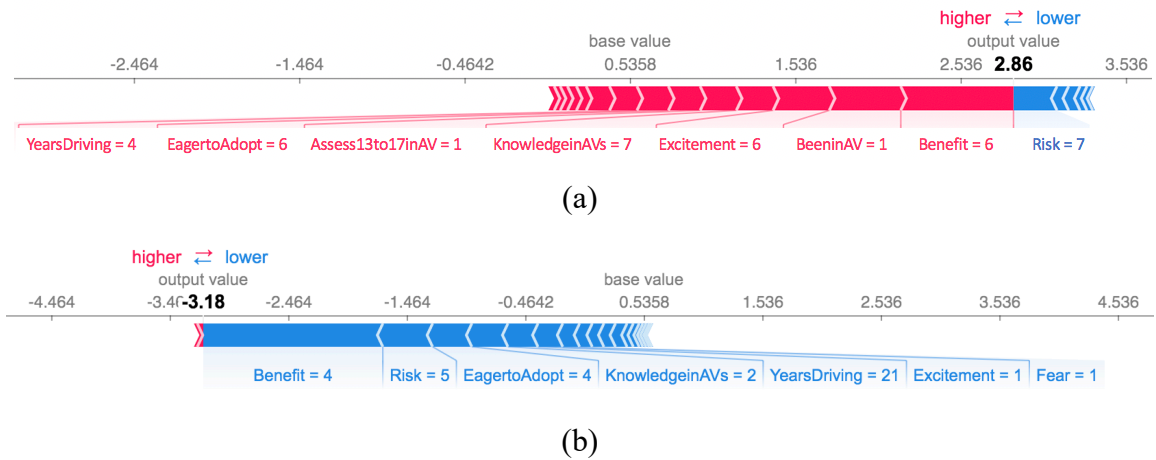

(a)

(b)

**Fig. 6.** SHAP individual explanations of trust prediction for randomly selected participants with (a) ground truth = trust and (b) ground truth = distrust.

# 6. Discussion

## 6.1. Predictability and Explainability

XGBoost is an efficient and easy to use algorithm for tabular data classification and delivers high performance and accuracy as compared to other algorithms (Chen and Guestrin, 2016). In this research, we used XGBoost to predict people's trust in AVs with superior performance. Compared to other machine learning models, XGBoost performed the best among various metrics, including accuracy, ROC_AUC, recall, and F1 measure (see Table 3). The model converged within 60 iterations in our experiment and proved to be a feasible solution to predict trust in AVs.

In order to improve the explainability of the XGBoost model, we used SHAP explainer which offers a high level of model interpretability (Lundberg and Lee, 2017). SHAP has a fast implementation for tree-based models (e.g., XGBoost), which overcomes the biggest barrier (i.e., slow computation) for adoption of Shapley values. On top of the advantage of fast implementation, SHAP provides another two advantages including global and local interpretability. The global interpretability is represented by the contribution of the SHAP values in the model predictive decision. It represents the negative and positive effects of the most important factors on the model prediction as shown in Fig. 3. Such global interpretability is similar to the feature effect plot in linear regression models. Furthermore, the model is able to show both main effects of individual predictor variables and interaction effects between two predictor variables on trust, indicating how they influence the prediction results as evidenced in Fig. 4, Fig. 5, and Table 4. As for the local interpretability, SHAP enables us to explain the prediction of each observation since each one gets its own set of SHAP values as illustrated in Fig. 6. With the local and global interpretability comes the power of SHAP in providing a high level of model explainability.

## 6.2. Important Factors in Predicting Trust

Compared to linear regression models, our method uncovered the factor importance in predicting trust using the SHAP feature importance plots and the SHAP summary plot as shown in Fig. 3. Among all the predictor variables, the Benefit factor ranked the most important and was positively correlated with trust, consistent with previous research (Choi and Ji, 2015; Bearth and Siegrist, 2016). Furthermore, we found an interaction effect between Benefit and BeeninAV (see Fig. 4(a)). Even when the participants perceived AVs with low benefits, their interaction with AVs could potentially improve their trust in them. This was consistent with Brell et al. (2019), which showed that the experience with AVs significantly increased the perception of the benefits in AVs.

The second most important factor was risk (Fig. 3). In line with prior studies (Numan, 1998; Kim et al., 2008; Pavlou, 2003), our results showed that an increase in risk led to a decrease in trust. Risk was found to interact with BeeninAV (Fig. 4(b)). When the participants viewed AVs to be risky, experience with AV could potentially improve their trust in AV. This was also in concordance to previous research (Brell et al., 2019), which showed a decrease in risk perception in AVs with the increase of experience in AVs. Therefore, it is important that automotive manufacturers give more chances for the public (especially for those who perceive AVs with no benefits or high risks) to test AVs in order to improve their trust in AVs.

While both the third and fourth most important factors, i.e., Excitement and KnowledgeinAVs were positively correlated with trust in AVs. Risk was found to interact with Excitement (Fig. 4(c)) and KnowledgeinAVs (Fig. 4(d)). When the participants were not very excited about manual driving, they tended to trust the AVs more if the risk was low. Silberg et al. (2013) found that people who were less passionate about driving were more likely to lean toward using AVs if it was safe. When the participants were excited about manual driving, they trusted the AV more even if the risk was higher. Such trust, however, could be a type of over-trust associated with strong emotions, such as excitement. For example, Dingus et al. (2016) argued that excited or angry drivers were more likely to take risky driving even in highly automated driving. An increase in KnowledgeinAVs increased participants' trust in AVs (Fig. 4(d)) which was consistent with previous studies such as (Khastgir et al., 2018). However, it seemed counter-intuitive that those who rated AVs as risky trusted AVs more than those who rated AVs as not risky when the participants scored high on knowledge in AVs. To investigate the obtained results, we found that the percentage of participants who scored high on both KnowledgeinAVs and Risk was 27.9. In addition, out of those participants, 77.5 % considered AVs as beneficial. Thus, this result might be explained by the finding that the degree of knowledge in AVs affected the perception of balance between the risks and trust in AVs as Schmidt (2004) argued that the more one knew about the risks in an automation system, the higher the chances to accept it. In other words, these participants believed that the risky situations associated with AVs might be avoided by a better understanding of how to deal with such situations, such as the takeover transition period in SAE Level 3 AVs (Zhou, Yang and Zhang, 2020; Na, Yang and Zhou, 2020). Moreover, the belief of the benefits brought from AVs might also make them trust AVs more.

The EagertoAdopt factor was ranked number 5, and an increase in eagerness to adopt a technology increased the chances of trusting AVs which was in line with previous research (Edmonds, 2019; Raue et al., 2019) (see Fig. 4(e)). We also found that Fear affected the impact of EagertoAdopt on trust—at a high level of eagerness to adopt a new technology, a low level of fear in manual driving increased the chances of trusting AVs. Fear, which is an important factor in technology adoption, was shown to shape judgements, choices, and perception of risks (Lerner and Keltner, 2001). According to Shoemaker (2018), fearless driving was associated with no fear of change, thus leading to an eagerness of technology adoption.

Other factors involved in the study were less important compared to the ones listed above. Although Assess13to17inAV was ranked number 6, it was surprising to see that Assess5inAV and Assess 6to12inAV were less important in predicting trust in AVs. Intuitive, without trust in AVs, a parent would not let his/her children be in an AV. However, in our survey, we did not specify if they were the participants' children. Further research is needed to address this issue. Gender, age (years of driving), and education level were also found to be less important. However, as seen in Fig. 4(f), we found that trust was shown to decrease with an increase in the number of driving years. Furthermore, Benefit affected the impact of DrivingYears on trust—for larger than 10 years and smaller than 40 years of driving experience, a high level of benefits increased trust in AVs. In line with previous research, old people showed more concerns about trusting AVs despite its benefits in maintaining their mobility (Schoettle and Sivak, 2016).

As a summary, the measured trust is based on dispositional trust and initial learned trust (see Hoff and Bashir, 2015). The dispositional trust shows participants' overall tendency without any context of AVs and the initial learned trust is dependent on their previous knowledge or past experience (e.g., news reports on AV accidents) prior to interacting with AVs. This is because the majority of the participants (i.e., 77.3%) had no chance to interact with AVs and there was no interaction between the participants and AVs during this study. However, the dispositional trust and the initial learned trust measured in our paper are the baseline to form people's trust in AVs. Prior to any interaction with AVs, people have an inherent level of dispositional trust which is one of the major factors that influences people's purchase or use of AVs. Individual differences, such as age, gender, educational levels, as well as their learned knowledge about and experience in AVs shaped their perceived risks in and benefits of AVs, which in terms influence their dispositional and initial learned trust. Between these two types of trust measured in the survey, we found that the variables related to dispositional trust were more important and predictive than those related to initial learned trust as shown in Fig. 3(a). Nevertheless, unlike previous studies, the most important contribution of this study was proposing a trust prediction model with explainability to understand participants' trust in AVs. Automotive manufacturers can potentially make use of the relationships between these important factors and their trust to improve acceptance and adoption of AVs by providing training, spreading the benefits of AVs, explaining the possible risks, improving the design of the system, and creating appropriate emotional responses to AVs.

**6.3. Limitations**

First, due to the cross-sectional study design, we cannot examine how people's opinions change over time. Therefore, we only measured participants' trust in AVs in a snapshot. Also, as the majority of the participants had little prior experience with AVs, the trust is primarily based on their dispositional and initial learned trust. Longitudinal studies are needed to understand the dynamic trust relationships between users and AVs when they have chances to interact with AVs over time (Ekman et al., 2018). Further research should also be conducted to assess participants' dispositional, situational, and learned trust (see Hoff and Bashir, 2015) at a finer granularity, by querying participants' trust in AVs over time (Ruijten et al. 2018). Second, it was difficult for us to make sure the superior quality of the survey data from AMT. In this research, we used various techniques to overcome that, including shorter surveys, removing invalid data by examining their survey completion time and data patterns. However, quality can be affected by the compensation rate (Buhrmester et al., 2011) and running the screening procedures mentioned above might not be enough to ensure a high quality of responses. Third, our survey was quantitative without any qualitative data to explain our prediction model. It would be also important to verify such explanations using qualitative data from the participants themselves with open-ended questions.

## 7. Conclusion

In this paper, we predicted dispositional and initial learned trust in AVs with high accuracy and explainability. We conducted an online survey to collect a range of variables that were related to participants' trust in AVs. The survey data were then used to train and test an XGBoost model. In order to explain the XGBoost prediction results, SHAP was used to identify the most important predictor variables, to examine main and interaction effects, and to illustrate individual explanation cases. Compared with previous trust predictions models, our proposed method combines the benefits of XGBoost and SHAP with good explainability and predictability of the trust model.

Menon, N. (2015). Consumer Perception and Anticipated Adoption of Autonomous Vehicle Technology: Results from Multi-Population Surveys. *Graduate Theses and Dissertations*. https://scholarcommons.usf.edu/etd/5992

Menon, N., Pinjari, A., Zhang, Y., & Zou, L. (2016, January 1). *Consumer Perception and Intended Adoption of Autonomous Vehicle Technology – Findings from a University Population Survey*. https://trid.trb.org/view/1394249.

Miller, D. J. E., & Perkins, L. (2010). *Development of Metrics for Trust in Automation* (p. 18). AIR FORCE RESEARCH LAB WRIGHT-PATTERSON AFB OH SENSORS DIRECTORATE. https://apps.dtic.mil/docs/citations/ADA525259

Miller, D., Johns, M., Mok, B., Gowda, N., Sirkin, D., Lee, K., & Ju, W. (2016). Behavioral Measurement of Trust in Automation: The Trust Fall. *Proceedings of the Human Factors and Ergonomics Society Annual Meeting*, *60*, 1849–1853. https://doi.org/10.1177/1541931213601422

NHTSA2010. (2010). *Traffic Safety Facts 2010 A Compilation of Motor Vehicle Crash Data from the Fatality Analysis Reporting System and the General Estimates System*. National Highway Traffic Safety Administration National Center for Statistics and Analysis U.S. Department of Transportation. https://crashstats.nhtsa.dot.gov/Api/Public/ViewPublication/811659

Numan, J. H. (1998). *Knowledge-based systems as companions: Trust, human computer interaction and complex systems*. Undefined. https://www.semanticscholar.org/paper/Knowledge-based-systems-as-companions%3A-Trust%2C-human-Numan/afb2b16ea898a8fd5ec603a38e69c1d742e75e35

Paden, B., Cap, M., Yong, S. Z., Yershov, D., & Frazzoli, E. (2016). A Survey of Motion Planning and Control Techniques for Self-driving Urban Vehicles. *ArXiv:1604.07446 [Cs]*. http://arxiv.org/abs/1604.07446

Parasuraman, R., & Miller, C. A. (2004). Trust and etiquette in high-criticality automated systems. *Communications of the ACM*, 51–55.

Parasuraman, R., & Riley, V. (1997). Humans and Automation: Use, Misuse, Disuse, Abuse. *Human Factors: The Journal of the Human Factors and Ergonomics Society*, *39*(2), 230–253. https://doi.org/10.1518/001872097778543886

Pavlou, P. A. (2003). *Consumer Acceptance of Electronic Commerce: Integrating Trust and Risk with the Technology Acceptance Model* (SSRN Scholarly Paper ID 2742286). Social Science Research Network. https://papers.ssrn.com/abstract=2742286

Peters, E., Västfjäll, D., Gärling, T., & Slovic, P. (2006). Affect and decision making: A "hot" topic. *Journal of Behavioral Decision Making*, *19*(2), 79–85. https://doi.org/10.1002/bdm.528

Rajaonah, B., Tricot, N., Anceaux, F., & Millot, P. (2008). The role of intervening variables in driver–ACC cooperation. *International Journal of Human-Computer Studies*, *66*(3), 185–197.

Raue, M., D'Ambrosio, L. A., Ward, C., Lee, C., Jacquillat, C., & Coughlin, J. F. (2019). The Influence of Feelings While Driving Regular Cars on the Perception and Acceptance of Self-Driving Cars: Feelings and Self-Driving Cars. *Risk Analysis*, *39*(2), 358–374. https://doi.org/10.1111/risa.13267
29